\documentclass[preprint,aps,showpacs,amsmath,prb]{revtex4}
\usepackage{graphicx}
\usepackage{float}
\newcommand{\y }{\'{\i}}
\newcommand{\be }{\begin{equation}}
\newcommand{\ee }{\end{equation}}

\begin{document}


\title{Effect of CO desorption and coadsorption with O on the phase diagram of a Ziff-Gulari-Barshad model for the catalytic oxidation of CO}

\author{G.~M.~Buend\y a}
\author{E.~Machado}
\affiliation{Physics Department, Universidad Sim\'on Bol\y var,\\
Apartado 89000, Caracas 1080, Venezuela}
\author{P.~A.~Rikvold}
\affiliation{ Department of Physics and\\
Center for Materials Research and Technology,\\
Florida State University, Tallahassee, FL 32306-4350, USA}

\date{\today}

\begin{abstract}

We study the effect of coadsorption of CO and O on a
Ziff-Gulari-Barshad (ZGB) model with CO desorption (ZGB-d) for the
reaction CO + O $\rightarrow$ CO$_2$ on a catalytic surface.  Coadsorption of 
CO at a surface site already occupied by an O is introduced by an 
Eley-Rideal-type mechanism that occurs with probability 
$p$, $0 \leq p \leq 1$. We find that, besides the well-known effect of 
eliminating the second-order phase transition between the reactive state 
and an O-poisoned state, the coadsorption step has a profound effect on 
the transition between the reactive state and the CO-poisoned state. 
The coexistence curve between these two states terminates at a critical 
value $k_c$ of the desorption rate $k$ which now depends on $p$. 
Our Monte Carlo simulations and finite-size scaling analysis indicate 
that $k_c$ decreases with increasing values of $p$. For $p=1$, there 
appears to be a sharp phase transition between the two states only for 
$k$ at (or near) zero.

\end{abstract}


\maketitle

\section{Introduction}
\label{sec:I}

Catalytic reactions on surfaces are very complex, and their comprehensive understanding is vital, both for its intrinsic scientific interest and for its immense technological applications.~\cite{bond87, chri94} The scientific interest is due to the emergence of a rich and complex variety of phenomena, including chaotic behavior, bistability, critical phenomena, out-of-equilibrium phase transitions, etc.~\cite{bond87,marro99}
Due to its crucial role in industrial applications, the oxidation of CO on a transition-metal catalyst is one of the most studied reactions. In a pioneering work, Ziff, Gulari and Barshad (ZGB)~\cite{ziff86} proposed a deceptively simple model to describe some kinetic aspects of the reaction CO + O $\rightarrow$ CO$_2$ on a catalytic
surface in terms of a single parameter $y$, which represents the
probability that the next molecule arriving at the surface is a
CO, i.e.,., it is proportional to the partial pressure of CO. The model
exhibits two kinetic phase transitions, one continuous, at $y=y_1$, between an oxygen-poisoned phase and a reactive one, and a discontinuous one, at 
 $y=y_2$, between a reactive phase and a CO poisoned one. Unfortunately, there are important aspects of this catalytic reaction that are not reproduced by this simple model. Transitions between
states of low and high CO coverage, $\theta_{\rm {CO}}$ (the fraction of surface sites
occupied by CO), have been observed experimentally.~\cite{ehsa89} At low
temperatures, as $y$ increases, there is a discontinuous drop in
the CO$_2$ production rate; while above a certain critical temperature
the discontinuity disappears, and the CO$_2$ production decreases
continuously. This type of behavior can be reproduced by including a CO 
desorption rate, $k$, the so-called ZGB-d model.\cite{kauk89,alba92,bros92} 
Another feature that is erroneously predicted by the original ZGB model is the continuous phase transition at $y_1 > 0$, since experiments show that the reaction rate increases as soon as the CO concentration departs from zero.~\cite{ehsa89} The real
system does not present an oxygen-poisoned state because oxygen does
not impede the adsorption of CO.~\cite{ertl90,kris92} Several authors have shown that this transition can be eliminated by adding to the Langmuir-Hinshelwood (LH) mechanism that defines the original ZGB model an  Eley-Rideal (ER) step that allows a reaction between CO molecules in the gas phase and adsorbed O atoms on the surface.~\cite{meak90, tamb94} However, there are no experimental data indicating that such a reaction between free CO and adsorbed oxygen occurs, but there is experimental evidence that suggests that under certain conditions, it is possible that a CO molecule can be weakly adsorbed at a site already occupied by an oxygen atom.~\cite{camp80} Experiments based on photoemission electron microscopy have shown that CO can be adsorbed on a saturated O surface.~\cite{huan02} This coadsorption is a prerequisite for the reaction to happen, and it can be simulated by an ER-type mechanism that occurs with a certain probability. Some previous work on the ZGB model with
  coadsorption indicates that the production rate of CO$_2$ increases linearly with $y$ (in the region of low $y$), and that the continuous phase transition from the reactive state to the O poisoned state disappears.~\cite{meak90, huad02, qais04, mukh09} In these works it was assumed that the addition of the ER-type step does not alter the nature of the transition at $y_2$, but simply shifts it to a lower value of $y_2$.

We believe that the effect of the ER-type step on the ZGB-model deserves a more detailed analysis. In this work we therefore explore what happens when a coadsorption mechanism is added to the ZGB model with CO desorption. For the ZGB-d model, there is a
distinction between the high and low CO-coverage phases only for desorption rates $k$
below some critical value $k_c$, while above $k_c$ the CO coverage
varies smoothly with $y$. Thus, the transition value $y_2$ becomes a
function of $k$, corresponding to a coexistence curve $y_{2}(k)$ that
terminates at the critical point $y_2(k_c)$.~\cite{tome93,ziff92,mach05} 
The value of $k_c$ has been estimated by several authors. An earlier work 
based on fractal scaling of the interface between phases indicates that 
the first-order phase transition disappears for 
$k > k_c \approx 0.039$.\cite{bros92}  More recent estimates based on the 
probability distribution for the CO coverage (histograms), the fourth-order 
cumulant, and finite-size scaling theory give the values 
$k_c=0.040$,\cite{tome93} and $k_c=0.039$,\cite{mach05} respectively.
In this work we are particularly interested in studying how the nature of the 
transition at $y_2$ changes when the ER-type mechanism is added 
with probability $p$ to the ZGB-d model. 
To the best of our knowledge, such an analysis has not been performed 
previously. Our study is based on kinetic Monte Carlo simulations and 
finite-size scaling analysis of the data. 

The rest of this paper is organized as follows. In Sec. II, we
define the model and describe the Monte Carlo simulation
techniques used. In Sec. III A, we present and discuss the numerical
results obtained for the case of $k>0$. In Sec. III B, we perform a 
finite-size scaling analysis for the particular case of the standard 
ZGB model with an Eley-Rideal step ($k=0, p=1$). 
Our conclusions are summarized in Sec. IV.

\section{Model and Simulation}
\label{sec:ModS}

The ZGB model with desorption and coadsorption is simulated on a square lattice of
linear size $L$ that represents the catalytic surface. A Monte Carlo
simulation generates a sequence of trials: CO or O$_2$ adsorption
with probability $1-k$ and CO desorption with probability $k$. For the adsorption a CO or O$_2$ molecule is selected
with probability $y$ and $1-y$ respectively.~\cite{ziff86,tome93}
These probabilities are the relative impingement rates of
both molecules and are proportional to their partial pressures.
The algorithm works in the following way. A site $i$ is selected
at random. For desorption, if $i$ is occupied by CO the site
is vacated; if not, the trial ends. For adsorption, if a CO
molecule is selected it can be adsorbed at an empty site $i$ if
none of its nearest neighbors are occupied by an O atom.
Otherwise, one of the neighbors occupied by O is selected at random and
removed from the surface, leaving $i$ and the selected neighbor
vacant. This move simulates the CO + O $\rightarrow$ CO$_2$ surface
reaction following the adsorption of CO. If the selected site is occupied by an O atom the CO is coadsorbed with probability $p$, and a CO$_2$ molecule is liberated leaving behind an empty site on the surface. This represents the ER reaction.
O$_2$ molecules can be adsorbed only if a pair of nearest-neighbor
sites are vacant. If the adsorbed molecule is selected to be
O$_2$, a nearest neighbor of $i$, $j$, is selected at random, and if it
is occupied the trial ends. If both $i$ and $j$ are empty, the
trial proceeds, and the O$_2$ molecule is adsorbed and dissociates
into two adsorbed O atoms. If none of the remaining neighbors of $i$ is
occupied by a CO molecule, the one O atom is located at $i$, and if
none of the neighbors of $j$ is occupied by a CO molecule, then the
other O is located at $j$. If any neighbors of $i$ are occupied by
a CO, then one is selected at random to react with the O at $i$ such
that both sites are vacated. The same
 reaction happens at site $j$ if any of its neighbors are filled
with a CO molecule. This process mimics the CO + O $\rightarrow$
CO$_2$ surface reaction following O$_2$ adsorption.
A schematic representation of this
algorithm is given by

\begin{eqnarray}
\text{CO(g)} + \text{S} & \rightarrow & \text{CO(a)} 
\nonumber \\
\text{O}_{2}(g) + 2\text{S} & \rightarrow & 2\text{O(a)} 
\nonumber\\
\text{CO(a)} + \text{O(a)} & \rightarrow & \text{CO}_2\text{(g)} + 2\text{S} 
\nonumber \\
\text{CO(a)}  & \rightarrow & \text{CO(g)} + \text{S} 
\nonumber \\
\text{CO(g)} + \text{O(a)} & \rightarrow & \text{O(a)}+\text{CO(c)} 
\nonumber \\
\text{CO(c)} + \text{O(a)} & \rightarrow & \text{CO}_2\text{(g)}+\text{S} 
\;
\nonumber
\end{eqnarray}
Here $S$ represents an empty site on the surface, $g$ means gas phase, $a$ means adsorbed, and $c$ means coadsorbed. 
The first three steps correspond to the LH mechanism, the fourth step to the CO desorption, and the last two steps to the ER-type mechanism, which occurs with probability $0\le p \le 1$. If $p=0$, we recover the ZGB-d model, if $p=1$, we have the ZGB-d model with an ER step. 
It is important to realize that the ER reactions  are essentially 
instantaneous, so that the CO cannot remain coadsorbed without reacting.

For our simulations we assume periodic boundary conditions. The
time unit is one Monte Carlo Step per Site, MCSS, in which each
site is visited once, on average. Averages are taken over $10^3$
independent simulation runs for each set of parameters.

\section{Results}
\label{sec:R}

Starting from an empty lattice, we wait until the system reaches a steady state at constant partial pressure $y$. We calculate the
fraction of sites occupied by CO molecules: the CO coverage,
$\theta_{\rm CO}$; the O coverage, $\theta_{\rm O}$, and the fraction of
empty sites, $\theta_{\rm S}$. $R_{\rm{CO}_2}$ is the rate of production of CO$_2$.

\subsection{Results for $k>0$}
\begin{figure}
\centering\includegraphics[scale=.6]{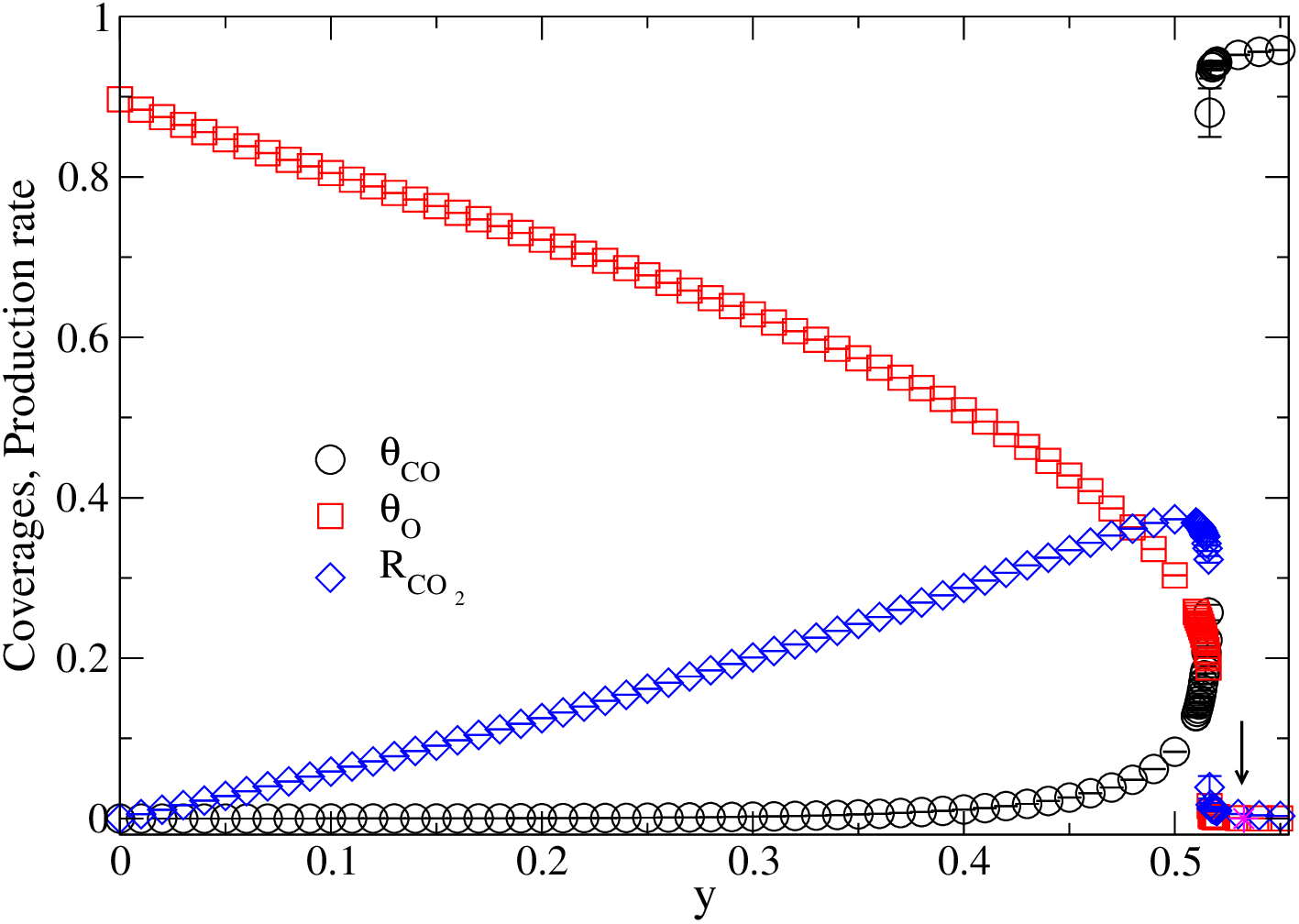}
\caption[]{Adsorption curves and CO$_2$ production rate for the ZGB-d model 
with an ER-type step for $p=0.5$, $k=0.02$, and $L=60$. 
Notice that in this case, the CO$_2$ 
production starts as soon as $y>0$, and it grows almost linearly until it 
reaches the transition at $y_2$, which occurs at a lower value than 
for $p=0$. Except right at $y_2$, the error bars are much smaller than 
the symbol size. The arrow points toward the value of $y_2$ for the 
pure ZGB-d model at the same value of $k$, 
$y_{2}(k=0.02, p=0)\approx 0.5327$.~\cite{mach05} It emphasizes the 
reduction in $y_2$ resulting from $p>0$.}
\label{cov1}
\end{figure}

\begin{figure}
\centering\includegraphics[scale=.5]{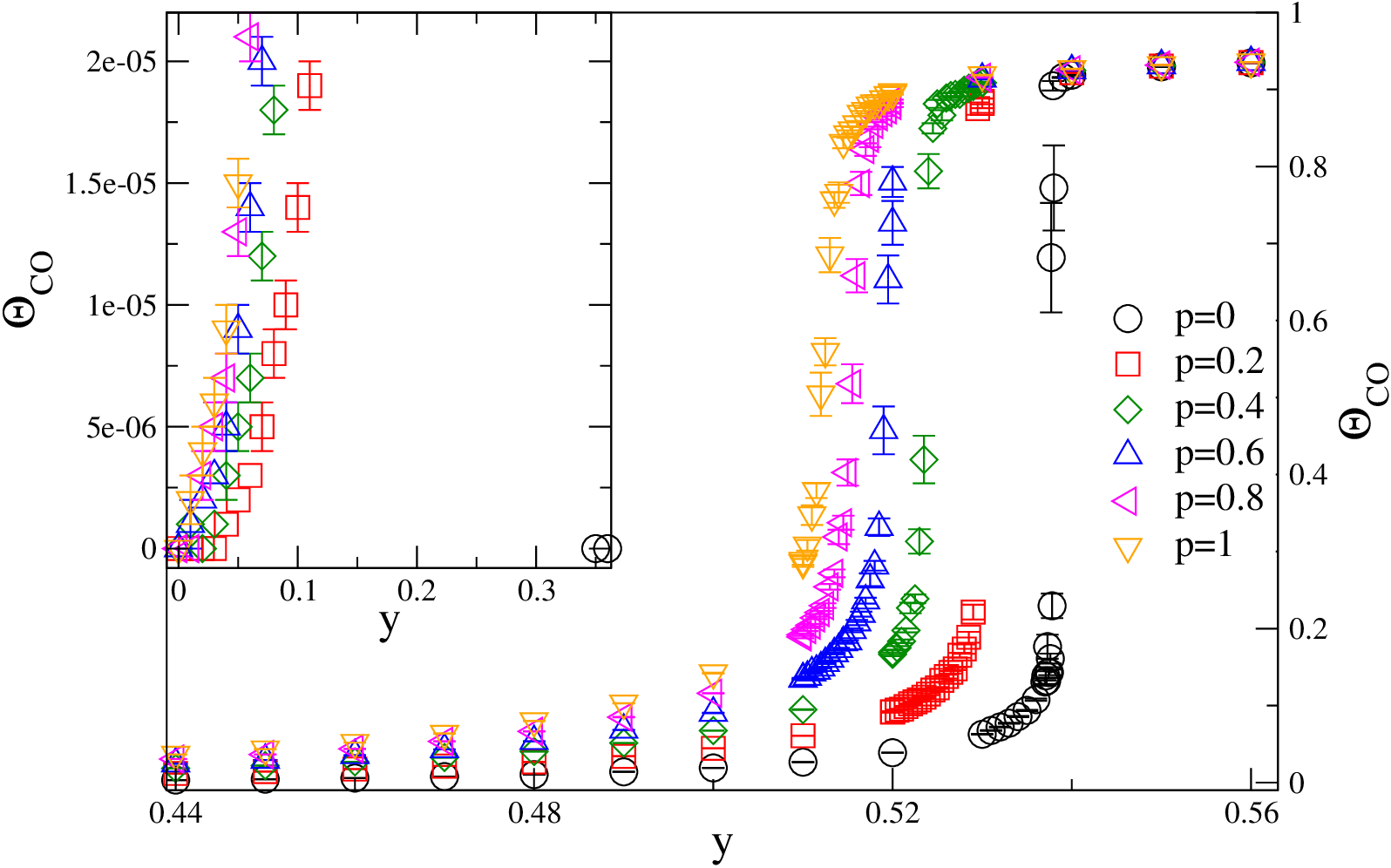}
\caption[]{CO coverage vs $y$ for desorption rate $k=0.03$ and several values of $p$. As $p$ increases, $y_2$ decreases. In the inset it can be seen that, except in the particular case $p=0$, in which the system has a phase transition at $y_1 \approx 0.388$, the CO coverage starts increasing as soon as $y>0$. $L=60$. }
\label{cov2}
\end{figure}

In Fig.~\ref{cov1}, we show the dependence of the coverages and the CO$_2$ production rate on the CO pressure, $y$, for the particular case of $k=0.02$ and $p=0.5$. It is immediately seen that the inclusion of the ER-type step ($p > 0$) has a dual effect: one is to eliminate the oxygen-saturated phase and therefore the continuous transition at $y_1$ in the ZGB model, and the other is that the transition to a CO-saturated surface occurs at a lower value of the CO pressure than for $p=0$. These two effects can be seen in more detail in Fig.~\ref{cov2}, where we plot the CO coverage vs $y$ for a fixed value of $k$ and different values of $p$. As $p$ increases, the second transition is shifted toward lower values of $y_2$, and the CO coverage starts to increase as soon as $y > 0$, except for the case $p=0$ (the ZGB-d model), in which it starts at 
$y_1 \approx 0.388$. A closer scrutiny of Fig.~\ref{cov2} also reveals that 
the discontinuity in $\theta_{\rm CO}$ at $y_2(k=0.03,p)$ 
appears to become smaller as $p$ increases, giving way to a continuous change 
for $p$ below approximately $0.6$.

Previous studies show that when CO desorption ($k>0$) is added to the original 
ZGB model, the ZGB-d model, the discontinuous transition ends at a critical 
value $k_c$ , such that in this case 
$y_{2}=y_2(k_c)$.~\cite{tome93,ziff92,mach05}. 
To the best of our knowledge, 
there have been no studies of how the nature of the 
transition changes when an ER-type step is added to the ZGB-d model, 
and this is the main topic we explore in this paper.

An efficient way to locate and classify phase transitions is to look at the 
fourth-order reduced cumulant of the order parameter. 
For $\theta_{\rm CO}$ it takes the form,~\cite{mach05,land00,bind84,challa86}
\begin{equation}
u_L = 1 - \frac{\mu_4}{3\mu_2^2} \;,
\end{equation}
where
\begin{equation}
\mu_n =  \langle (\theta_\text{CO}-\langle
\theta_\text{CO}\rangle )^n\rangle=
\int_0^1 (\theta_\text{CO}-\langle \theta_\text{CO}\rangle)^n
P(\theta_\text{CO}) d\theta_\text{CO}
\end{equation}
is the $n$th central moment of the CO coverage,  
and $P(\theta_{\rm CO})$ is the probability distribution for $\theta_{\rm CO}$. This probability distribution is obtained by counting the number of times, $N_i$, that the
coverage falls in the intervals $[0, \Delta), [\Delta, 2\Delta),
...,[1-\Delta, 1]$ ($\Delta = 0.01$), such that $\sum_i N_i = N$
is the total number of MCSS. Then, the probability, $P_i$, that $\theta_{\rm CO}$ has a value
in the interval $[(i-1)\Delta, i\Delta)$ is
\begin{equation}
P_i=\frac{N_i}{N\Delta} \;,
\end{equation}
such that
\begin{equation}
\int_0^1 P(\theta_\text{CO}) d\theta_\text{CO} \approx \sum_i P_i\Delta=1 \;.
\end{equation} 

\begin{figure}
 \centering\includegraphics[scale=.4]{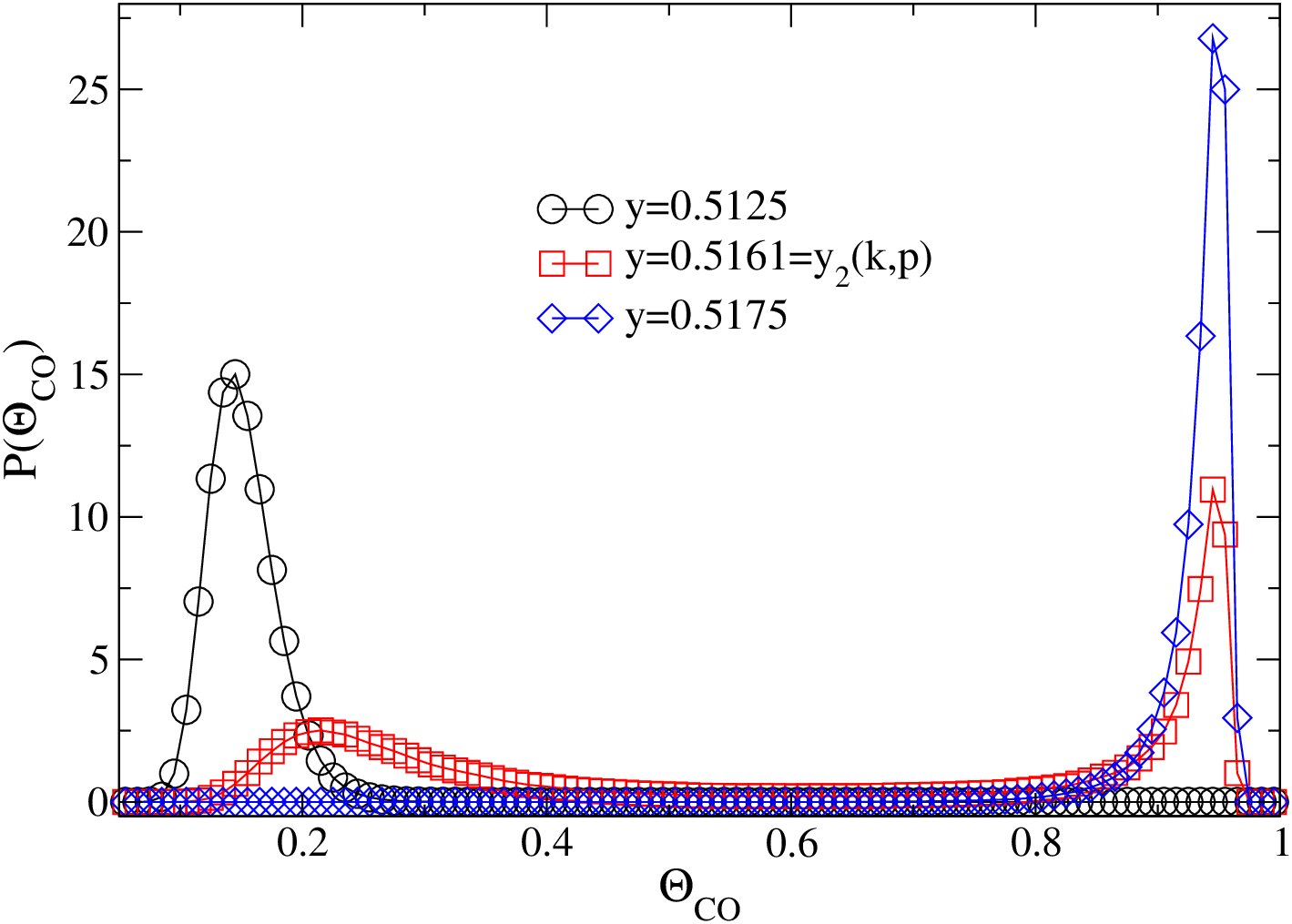}
\caption[]{Order-parameter probability distribution $P(\theta_\text{CO})$ for fixed $k=0.02$ and $p=0.5$, below ($y=0.5125$, circles), and above  ($y=0.5175$, diamonds) the transition at $y_2$, where the distributions are unimodal and remain in the low and high coverage phases respectively. At the transition point, $y_{2}(k=0.02,p=0.5) \approx 0.5161$ (squares), the histogram is bimodal and the areas under the peaks are equal, suggesting the coexistence of two different phases. $L=60$.}
\label{hist}
\end{figure}

In Fig.~\ref{hist} we show $P(\theta_\text{CO})$ for fixed values of $k$ and $p$, and values of $y$ in different regions: below and above $y_2$, where the histograms are unimodal, indicating that the system consists of one single phase, and at $y_2$ where the histograms are bimodal, signaling the existence of two distinct phases. At the coexistence point $y_2(k,p)$, the areas under both peaks are equal.~\cite{borg90, land00} A finite-size scaling analysis would have to be performed to ascertain if the transition is continuous or discontinuous for any particular values of $k$ and $p$. Such a study is planned for future work.

\begin{figure}
\centering\includegraphics[scale=.45]{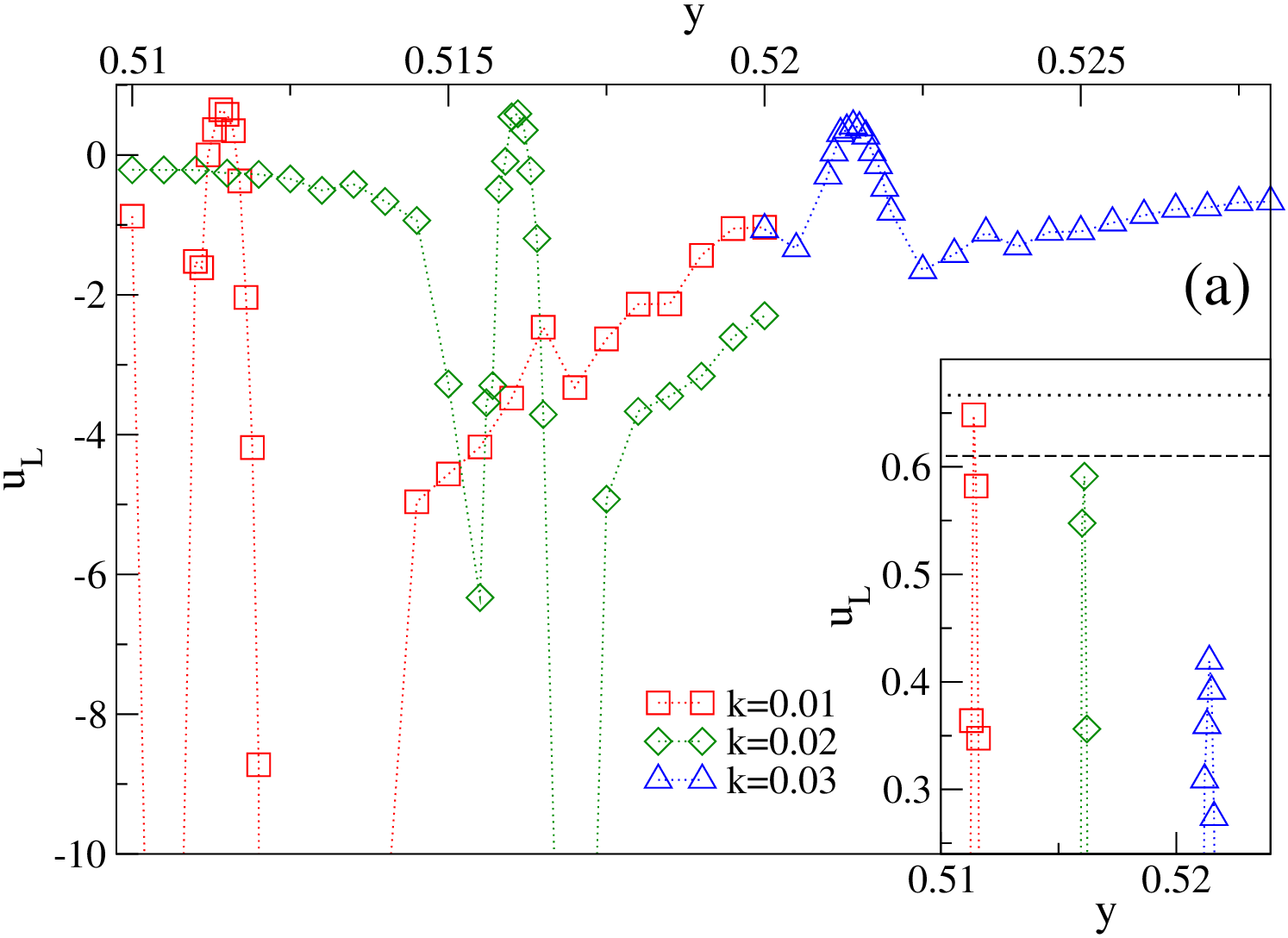}\\
\vspace{1.1truecm}
\centering\includegraphics[scale=.45]{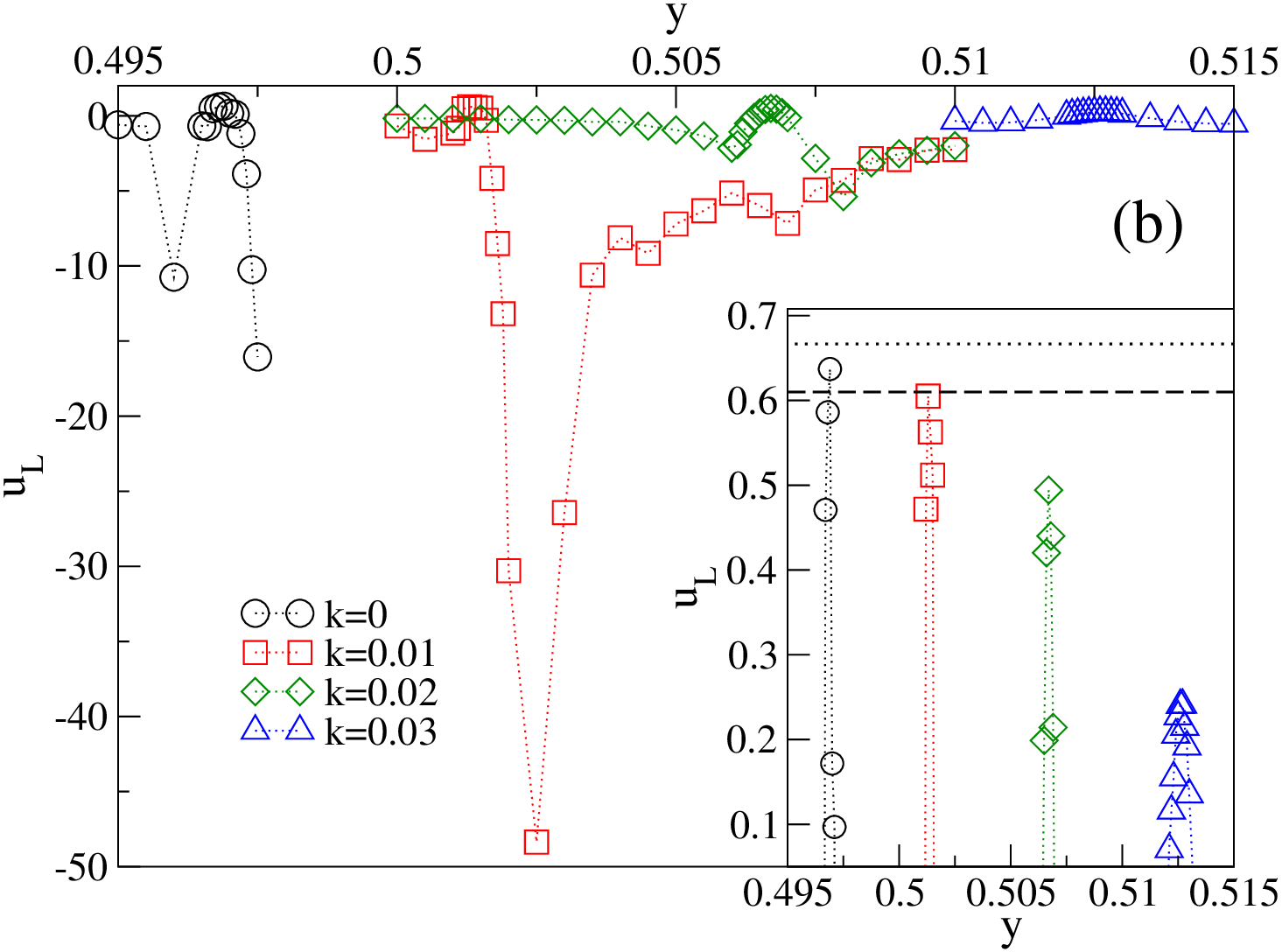}
\caption[]{Fourth-order cumulant $u_L$ vs $y$ for (a) $p=0.5$ and (b) $p=1$. 
The minima of $u_L$ correspond to the crossovers between unimodal and bimodal 
distributions. The maximum between these minima gives the coexistence 
point, $y_2(k,p,L)$. The dotted horizontal lines in the insets correspond 
to $u_L=2/3$, and the dashed horizontal lines to $u_{Ising} \approx 0.61$. 
$L=60$.
For $p=0.5$ (a), the maximum occurs at the same value of $y$ (0.516), for which 
the areas under the two peaks of the order-parameter distribution in 
Fig.~\protect\ref{hist} are equal.}
\label{uL_vs_y}
\end{figure}

The equal-area bimodal distribution corresponding to coexistence
yields a positive maximum for the cumulant $u_L$ vs $y$,
flanked on either side by negative minima and approaching zero far away from
the transition.
The maxima of $u_L$ define the
$L$-dependent coexistence line, $y_2(k,L,p)$. In Fig.~\ref{uL_vs_y}
we show the dependence of $u_L$ on $y$ for several values of of 
$k$ and $p=0.5$ (Fig.~\ref{uL_vs_y}(a)), and $p=1.0$ (Fig.~\ref{uL_vs_y}(b)). 
In the proximity of a first-order phase transition, the maximum value of the 
fourth-order cumulant (when $L \rightarrow \infty$) takes the value 
of $2/3$,\cite{bind81} while 
for a continuous transition the maximum takes a lower value. For the Ising 
universality class, the maximum value of $u_L$ at the critical point is 
approximately  0.61.~\cite{kami93} Our results strongly suggest that when 
the ER-type step is added ($p>0$) the coexistence curve ends at some 
value of $k_c$ that decreases when $p$ increases, i.e., now $y_2(k_{c},p)$. 
Figure \ref{uL_vs_y}(a) indicates that when $p=0.5$, the first-order phase 
transition seems to terminate at some value of $k$ below 0.02, 
i.e., $0.01 < k_{c}(p=0.5) < 0.02$, while Fig.~\ref{uL_vs_y}(b) indicates 
that even for the well-known ZGB model ($k=0$, no desorption) 
with $p=1$,~\cite{meak90} it is not clear that the transition remains 
first-order. In the following we analyze this particular case in further detail.

\subsection{Results for $k=0, p=1$}

The standard ZGB model with the addition of an ER step has been analyzed by several authors,~\cite{meak90,tamb94,huad02,qais04, mukh09} but as far as we know the nature of the transition at $y_2$ has not been studied in detail, and it has been assumed that, besides destroying the continuous transition at $y_1$, the effect of the extra step is simply to shift the first-order transition to a lower value of $y_2$.
Our analysis of the fourth-order cumulant suggests that $k_c$ is depressed by increasing $p$, and in this section we perform finite-size scaling analysis to further investigate this point.

\begin{figure}
\centering\includegraphics[scale=.6]{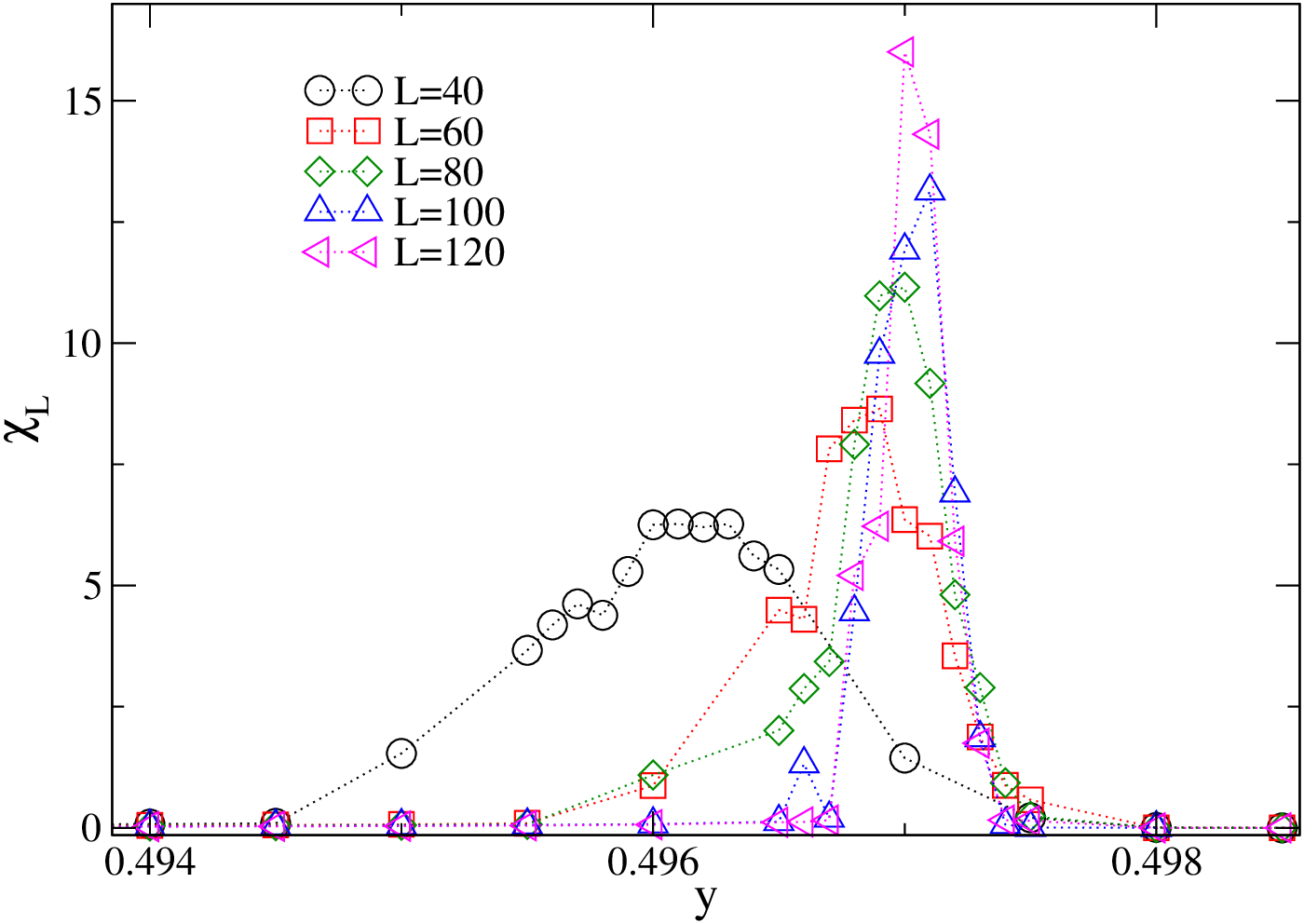}
\caption{ Fluctuation of the order parameter, $\chi_L$, vs $y$ for $k=0$, $p=1$ for different system sizes $L$.}
\label{chi}
\end{figure}

\begin{figure}
\centering\includegraphics[scale=.4]{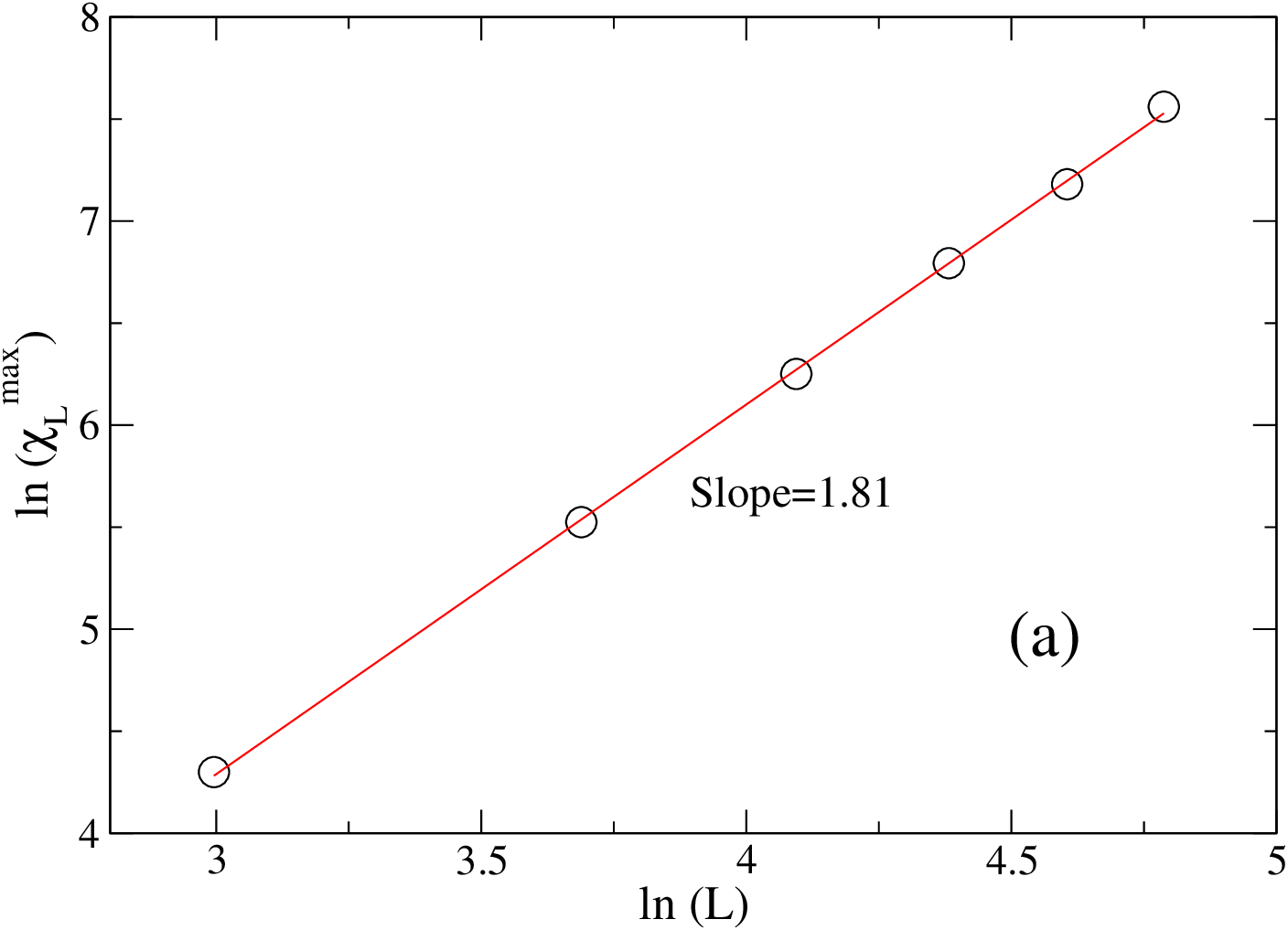} \\
\vspace{1.1truecm}
\centering\includegraphics[scale=.4]{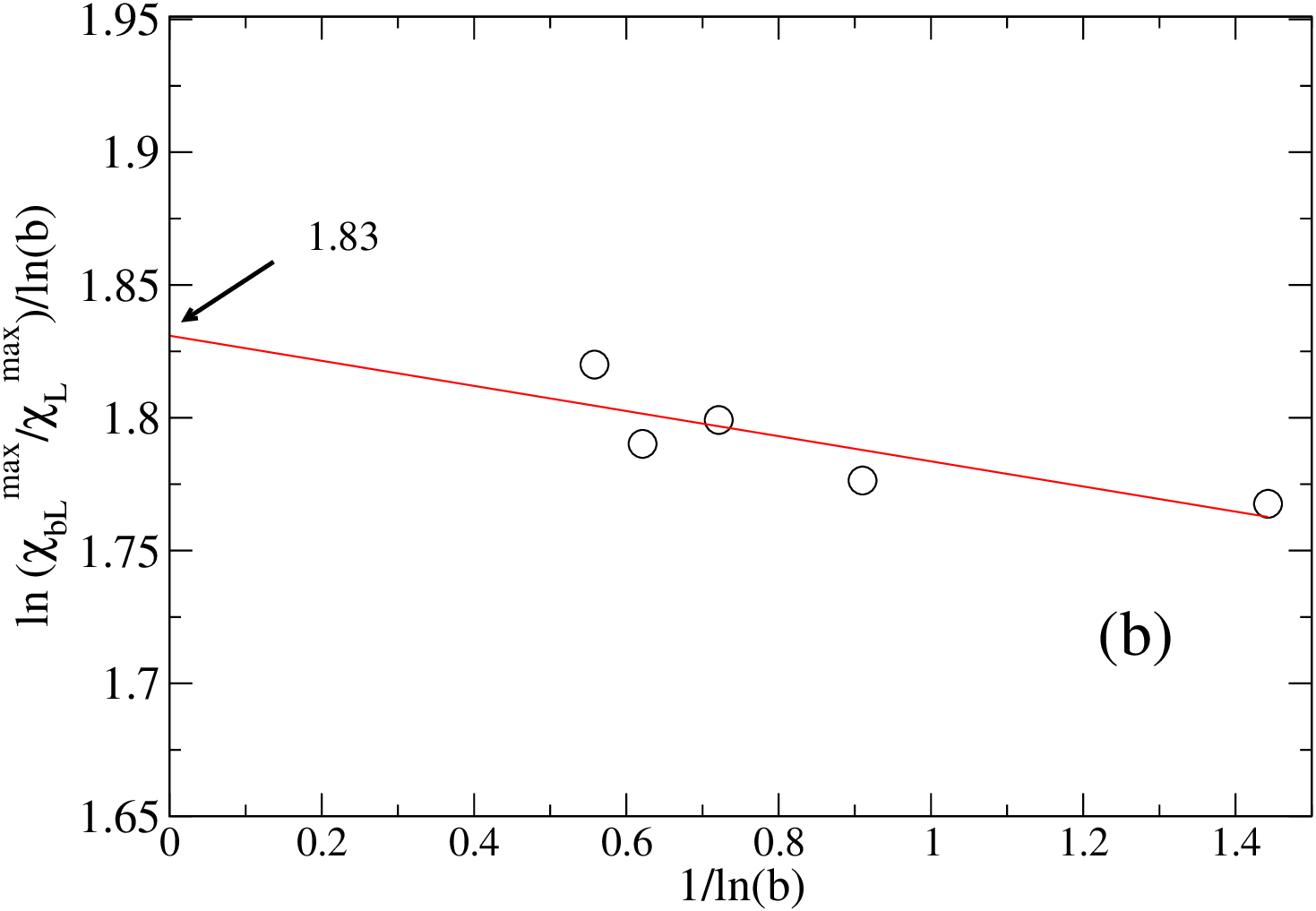}
\caption{Scaling behavior of $\chi_{L}^\text{max}$ for the ZGB model with an ER step ($k=0, p=1$). (a) ln($\chi_{L}^\text{max}$) vs ln(L) and (b) ln ($\chi_{bL}^\text{max}$/$\chi_{L}^\text{max})$/ln (b) vs 1/ln(b) with L=20. The straight lines are the best linear fits to the data and give $\chi_{L}^\text{max}\sim L^{d^{\prime}}$. From (a) $d^{\prime}=1.81\pm0.02$ and from (b) $d^{\prime}=1.83\pm0.02$. }
\label{fss}
\end{figure}

The fluctuations of $\theta_\text{CO}$ can be calculated in the standard way as
\begin{equation}
\chi_L=L^{2}(\langle \theta_\text{CO}^2 \rangle_{L}-\langle \theta_\text{CO}\rangle_{L}^{2}),
\end{equation}
where
\begin{equation}
 \langle \theta_\text{CO}^n \rangle_{L}=\int_0^1\theta_\text{CO}^nP(\theta_\text{CO})d\theta_\text{CO}.
\end{equation}
The peak positions of $\chi_L$ approach the infinite-system transition point with increasing $L$. In Fig.~\ref{chi} we show $\chi_L$ as a function of $y$ for $k=0$ and $p=1$ for several values of
$L$. As expected, the peaks of the curves shift and increase in height with increasing $L$.
It is well known that at a first-order equilibrium phase transition the 
order parameter fluctuations scale as $\chi_L \sim L^d$, where $d$ is the 
spatial dimension of the system (here $d=2$).~\cite{nien75,fish82,land00}
Previously we have successfully used the same scaling relation to identify the 
{\it nonequilibrium\/} first-order transition in the ZGB-d model.\cite{mach05} 
In Fig.~\ref{fss}(a)
we plot ln($\chi_{L}^{\rm max}$) vs ln($L$). The linear fit indicates a power-law scaling, such that $\chi_{L}^\text{max} \sim L^{d^\prime}$, where $d^{\prime}=1.81\pm0.02$. We also measure the power-law exponent by a different method, which has some advantage
in eliminating the effects of a nonsingular background term
(as in $\chi_{L} = f + gL^{d^{\prime}}$ with $f$ and $g$ constants). We write 
\begin{equation}
\ln\left[\frac{\chi_{bL}^\text{max}}{\chi_L^\text{max}}\right]/\ln b=
d^{\prime} + {\cal O}(1/\ln b)
\label{eq:bLX}
\end{equation}
with $L$ fixed at a relatively small value (here, $L=20$), and $b>1$.
For large $L$ and $b$,
the correction term is proportional to $f/(g \ln b)$, so that the exponent
can be estimated by plotting the left-hand-side of Eq.~(\ref{eq:bLX}) vs
$1/\ln b$ and extrapolating to $1/\ln b = 0$, as in
Fig.~\ref{fss}(b). The resulting estimate is
$d^{\prime}=1.83\pm 0.02$. These results are relatively far from the expected value $d^{\prime}=2$ for a first-order phase transition, in clear contrast with the exponent calculated by the same techniques for $p=0$, in which case we found that $d^{\prime}\approx2$ for the transition at $y_2(k<k_{c},p=0)$.~\cite{mach05}
This analysis provides further support to our previous suggestion, based on the fourth-order cumulant, that the ZGB model with an ER step ($p=1$) does not have a phase transition but only a smooth crossover for any non-zero value of $k$.

\section{Conclusions}
\begin{figure}
\centering\includegraphics[scale=.6]{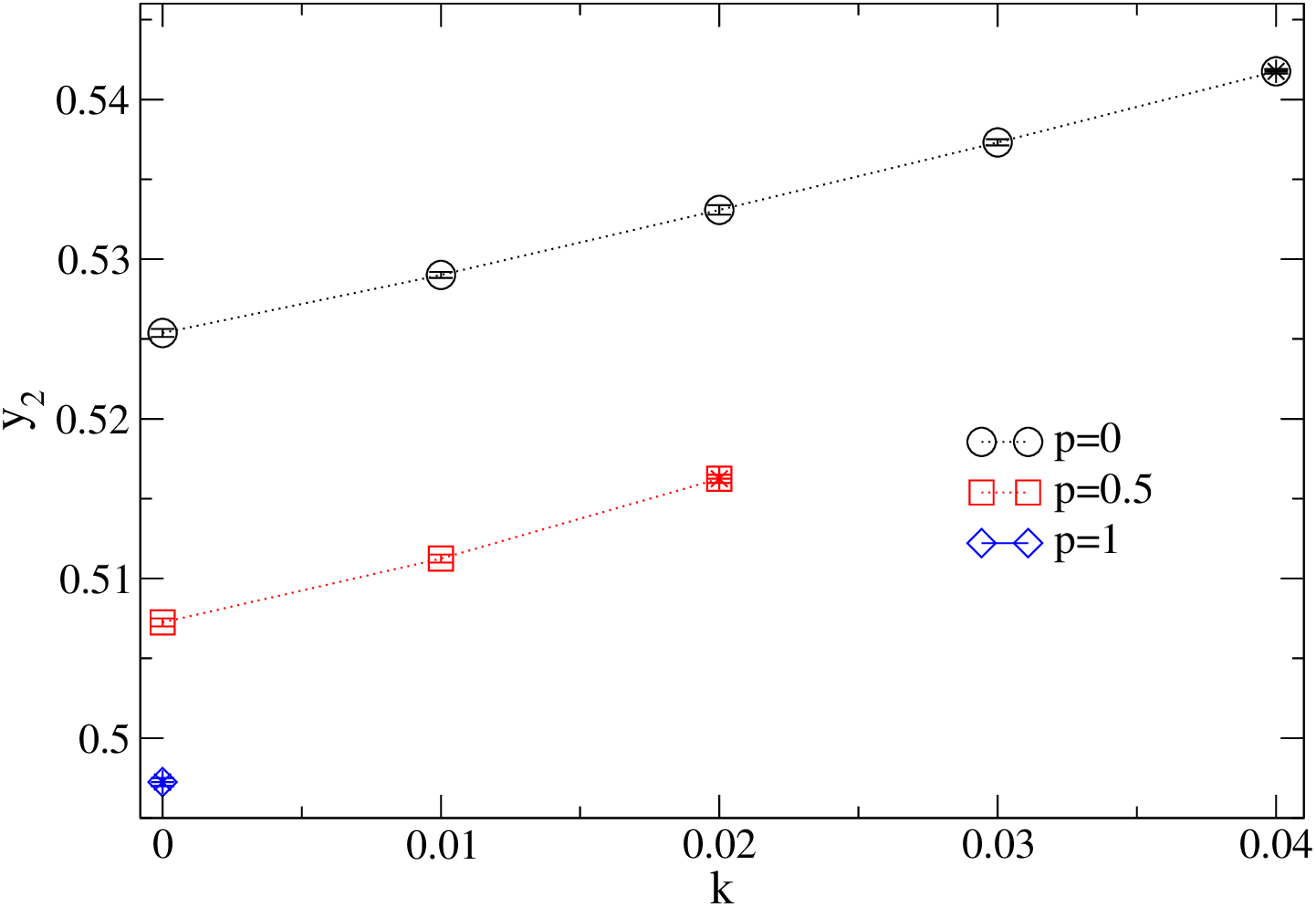}
\caption{Some points on the coexistence curve $y_2$ shown vs $k$ for different values of $p$. In each case the symbol with the asterisk indicates our highest estimate of $k_c(p)$, where the coexistence curve terminates at a critical point. For $p=1$, $k_c$ appears to be at (or close to) zero.}
\label{coex}
\end{figure}

In this paper we have presented a study of the ZGB-d model in which an ER-type step is added
($0 \le p \le 1$). The extra step can be thought of as a way to include a 
surface interaction between CO and O coadsorbed at the same site. 
We found that, besides eliminating the unphysical oxygen-poisoned state and 
therefore the second-order transition of the original ZGB model, the 
inclusion of the ER-type step has a more profound effect than the mere 
shifting of the first-order transition at $y_2$ toward lower values, 
which was previously reported in the 
literature.~\cite{meak90,tamb94,huad02} The addition of the ER-type step 
changes the nature of the transition between the CO$_2$ producing phase and 
the non-productive CO-poisoned phase, in such a way that now the coexistence 
curves terminate at a value of $k_c$ that decreases with $p$. Even for the 
well-studied case of the ZGB with an ER mechanism ($p=1$), our results indicate 
that it is not legitimate to speak of a phase transition at $y_2$ for any 
non-zero $k$. These results are summarized in Fig.~\ref{coex}, where 
we show a rough estimate of the coexistence curves indicating for some 
values of $p$ the upper value of $k$ for which there could still exist a 
first-order phase transition.
We hope our results will be of use for the analysis of experiments on
heterogeneous CO oxidation.

\section*{Acknowledgments}
G.M.B is grateful for the hospitality of the Physics Department and the Center for Materials Research and Technology at Florida State University. P.A.R acknowledges support by U.S. National Science Foundation Grant No. DMR-0802288.

\end{document}